# Selective thermal evolution of native oxide layer in Nb and Nb₃Sn-coated SRF grade Nb: An *in-situ* angular XPS study.

Arely Cano[1], Grigory V. Eremeev[1]*, Juan R. Zuazo[2], Jaeyel Lee[1], Bing Luo[3], Martina Martinello[1,4], Alexander Romanenko[1], Sam Posen[1]*

[1]Fermi National Accelerator Laboratory, Batavia, Illinois, 60510, USA.
[2]Spanish CRG beamline at the European Synchrotron Radiation Facility, B.P. 220, F-38043 Grenoble, France
[3]Characterization Facility, University of Minnesota, 12 Shepherd Labs, 100 Union St. SE, Minneapolis, MN, 55455, USA
[4]SLAC National Accelerator Laboratory, Menlo Park, CA, 94025, USA

**ABSTRACT**

This contribution discusses the results of an *in-situ* angular XPS study on the thermal evolution of the native oxide layer on Nb₃Sn and pure Nb. XPS data were recorded with conventional spectrometers using an AlK$_\alpha$ X-ray source for spectra collected up to 600 ºC, and a MgK$\alpha$ X-rays source for temperatures above 600 ºC. The effect of the thickness, composition, and thermal stability of that oxide layer is relevant to understanding of the functional properties of superconducting radiofrequency (SRF) cavities used in particle accelerators. There is consensus that oxide plays a role in the surface resistance (Rs). The focus of this study is Nb₃Sn, which is a promising material that is used in the manufacturing of superconducting radiofrequency (SRF) cavities as well as in quantum sensing, and pure Nb, which was included in the study for comparison. The thermal evolution of the oxide layer in these two materials is found to be quite different, which is ascribed to the influence of the Sn atom on the reactivity of the Nb atom in Nb₃Sn films. Nb and Sn atoms in this intermetallic solid have different electronegativity, and the Sn atom can reduce electron density around neighboring Nb atoms in the solid, thus reducing their reactivity for oxygen. This is shown in the thickness, composition, and thermal stability of the oxide layer formed on Nb₃Sn. The XPS spectra were complemented by grazing incident XRD patterns collected using the ESRF synchrotron radiation facility. The results discussed herein shed light on oxide evolution in Nb₃Sn compound and guide its processing for potential applications of the Nb₃Sn-based SRF cavities in accelerators and other superconducting devices.

**HIGHLIGHTS:** XPS study of native surface oxide layer in Nb₃Sn and pure Nb samples; Thermally induced changes in the surface oxide layer of Nb₃Sn and pure Nb; Phase composition of the surface oxide layer of Nb₃Sn and pure Nb from combined grazing incident XRD and XPS data; Thickness of the surface oxide layer in Nb₃Sn and pure Nb; Nature of the observed differences for the surface oxide layer in Nb₃Sn and pure Nb.

**GRAPHICAL ABSTRACT**

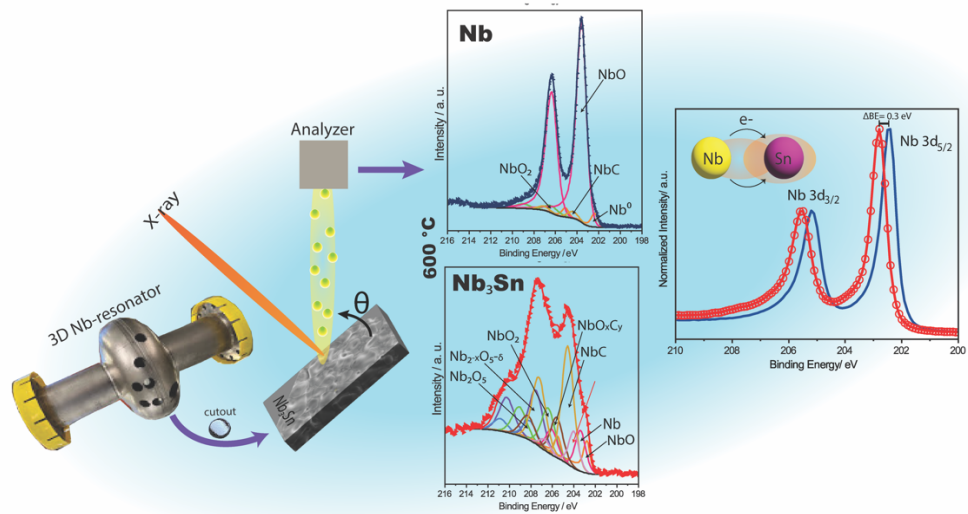

**KEYWORDS.** Reactivity of Nb in Nb₃Sn. XPS study of the surface oxide layer in Nb₃Sn; Features for the Nb₃Sn surface oxide layer. Thermal stability of the Nb₃Sn surface oxide layer. Changes in the Nb₃Sn oxide layer on heating.

**INTRODUCTION**

The small-scale accelerators using Nb₃Sn technology are in the prototyping stage thanks to world-wide advances in multi-disciplinary areas [1, 2] such as cryogenics, materials science, surface engineering, etc. Nb₃Sn is leading candidate as an alternative superconductor in RF technologies mainly due to the potential accelerating gradients of about 100 MV/m [3], twice the theoretical limit of pure Nb, and the possibility to operate at higher cryogenic temperatures than those required for Nb. Indeed, the residual surface resistance of Nb₃Sn is particularly low, in the range of n$\Omega$ [4]. Similar values have been reported for pure Nb cavities with outstanding gradients [5, 6], and even lower values were measured in Nb cavities, that were heat treated to partially dissolve surface oxide layer. This oxide dissolution is the surface-engineering technique that reduces the residual surface resistance at low accelerating fields through the dissolution of $Nb_2O_5$ [7]. Several studies have found evidence that the decrease of the quality factor at low fields could be related to microwave losses in $Nb_2O_5$ layer [8, 9, 10]. Niobium surface oxide on polycrystalline Nb is mainly composed of amorphous $Nb_2O_5$ and $Nb_xO_y$ sub-oxides, with an estimated thickness close to ~5 nm [11]. The nature and thickness of the oxide layer have been widely characterized through routine electron spectroscopies such as Auger Spectroscopy and XPS [6, 12, 13, 14, 15], and most recently using techniques with atomic resolution such as TEM and STEM [16, 17, 18]. As mentioned above, electron spectroscopies are powerful surface-specific analytical techniques useful in the study of the electronic structure of amorphous and recrystallized surface oxides. To our knowledge, the protective surface layer of Nb₃Sn remains less studied. Compared with metallic Nb, Nb atoms in Nb₃Sn, must be less reactive to the oxidation process because they are involved in a bonding interaction with the Sn atom, which is more electronegative than Nb. Such an interaction lowers electron density on the Nb atom, compared with the one for metallic Nb, and reduces reactivity for oxygen. This reduction in electronegativity of Nb could explain the slow growth of the native oxide layer on the surface of Nb₃Sn, resulting in a certain degree of the ordering for oxide phases, and higher thermal stability, which has implications for the thickness and stability of the oxide layer. In this study, we studied the surface structure, the chemical composition, and native oxide thickness together with the respective changes upon thermal treatments in under Ultra High Vacuum (UHV) conditions. Continuous XPS data recording during heating was used extract information on the surface oxide evolution. The initial experiments were carried out to evaluate the chemical composition by cross-section EDS maps, in combination with XPS spectra and oxide-layer thickness calculation from the angle variation approach (ARXPS) at room temperature. Then, *in-situ* angular XPS was used to follow the selective natural oxide layer evolution in Nb₃Sn and pure Nb in high temperatures, > 600 ºC. Finally, in the last section, we discuss the formation of NbC and the degassing of carbonaceous contaminants at sample temperatures below 300 ºC.

**EXPERIMENTAL SECTION**

*EP-Nb sample preparation*: The sample labelled as EP-Nb has a pill shape (Diameter =1 cm and thickness= 3 mm). It was extracted from a 1.3 GHz SRF Nb cavity labeled as TE1AES008. The RF baseline treatment was applied to this cavity, and it is well-described elsewhere [6, 7, 19]. Each sample was carefully extracted from the cavity following the established cut-out protocol. This protocol [20] was applied to avoid sample heating and further introduction of contaminant particles.

*Nb₃Sn sample preparation:* From here, we are going to refer to Nb₃Sn-coated SRF grade Nb samples such as Nb₃Sn samples. Nb₃Sn coating protocol followed in this study has been previously reported elsewhere [2, 21]. In short summary, niobium surface is anodized at 30-50 V, which results in 60-100 nm growth of Nb-oxides. After the anodization process, the cavity surface color turns from shinny mirror-like gray into metallic blue, see Figure S1. The cavity and samples are then introduced to the furnace under ultra-high vacuum, $10^{-7}$ mBar for a period of 24 hours at 200 ºC. After this low temperature heating to improve vacuum, the furnace is heated to 500 ºC and held at this temperature for 3 hrs. The last heating step is carried out close to 1200 ºC for several hours [2]. After the cooldown to room temperature, the assembled samples are removed and washed several times with manual high-rise pressure (HPR) and dried in a controlled atmosphere.

*Grazing Incident X-ray Diffraction:* The Grazing Incident X-ray Diffraction test was run at the BM25 Spline beam station of the European Research Synchrotron facilities (ESRF). The XRD experimental set-up is composed of a six-circle diffractometer, a 2x2 Maxipix detector and a large area CCD detector. The measurements were acquired to a maximum energy of 15 keV to avoid the Nb K$\alpha$ fluorescence signal (Nb: 20-25 keV) [22]. The calculated depth penetration for $\lambda$= 0.826 Å and 1º degree of incidence angle is about ~1 $\mu$m, using Parratt's equation [23]. The patterns were indexed with the Dicvol method in conjunction with a crystalline phase identification using Match! software loaded with the ICSD database. The LeBail fitting and refined profile were carried through the FullProof package [24].

*TEM sample preparation:* Details of the surface of $Nb_3Sn$ coating were systematically evaluated employing TEM. A thin TEM foil was prepared using a 600i Nanolab Helios focused ion-beam (FIB) and the TEM samples were thinned using a 30 kV $Ga^+$ ion-beam at 87 pA. The samples were then fine-polished utilizing 5 kV $Ga^+$ ions at 28 pA, to remove the damaged surface layers of the TEM foils. HR-STEM imaging was performed using an aberration-corrected JEOL ARM 200 and energy dispersive X-ray spectroscopy (EDS) mapping of Nb, Sn, O, which were collected to analyze the compositional changes in the surface oxide of $Nb_3Sn$ coatings.

*XPS measurements.* The in-situ XPS measurements were carried out with a PHI 5000 VersaProbe III spectrometer from ULVAC PHI, Inc. equipped with a monochromatic Al K$\alpha$ beam and Ar cluster ion source (GCIS). The chosen spot size was 100 μm; the spectra were recorded with a pass energy of 26 eV and 0.05 as the step size. The samples were measured as received and after 10 kV Ar cluster ion cleaning, which removes about 7 nm/min for organic contaminants calibrated with a PMMA film. It is estimated that when sputtering inorganics using this GCIB condition, the sputter rate is less than 1 Å /min. The Angle-dependent XPS measurements, ARXPS, were acquired by tilting the sample, conventional approach, with emission angles, $\theta_e$: 80º, 60º, 45º, 35º, 25º, 15º set to the normal surface.

*In-situ XPS measurements:* The $Nb_3Sn$ witness sample was mounted on the support of the sample holder; on top of the sample a thin metallic sheet with an area opening was placed with fixing screws. The sample was kept into the load-lock until the vacuum pressure reached $10^{-6}$ mBar. Then, the sample was transferred to the analysis chamber, where the vacuum pressure is steady at $10^{-9}$ mBar. The sample was tilted to 34º, near the most instrumental grazing configuration. The heating rate was set at 5 ºC/min. At every 100 ºC, from 26 to 600 º C, the temperature was fixed and maintain for 1 hr., followed by the recording of the high-resolution core-level spectra of C 1s, O 1s, Nb 3d, Sn 3d and Sn MNN. The same workflow was applied to the EP-Nb sample. During the heating process of EP-Nb sample, a drop in the vacuum pressure to $10^{-4}$ mBar was recorded above 145 ºC. Once the vacuum pressure was restored, the heating rate was set to 2

°C/min from 145 to 200 ºC, and the spectra registration carried out at every 10 ºC. After this critical range of temperatures, the heating rate was set back to 5 ºC/min, and the spectra was recorded every 100 ºC after 1 hr. at the temperature. The measurements at high-T for $Nb_3Sn$ samples (> 600 ºC) were performed in another XPS spectrometer mounted at the Spline Spanish beamline station at ESRF, France. The spectra were recorded using a non-monochromatic Mg K$\alpha$ source and an incident angle of 15º. The same annealing procedure was applied.

*XPS data analysis.* For binding energy calibration, Au $4f_{7/2}$= 84.0 eV was used was a reference. The .spe (PHI Multipack) files were manually converted to ISO files using CASAXPS software [25]. This software was used for the curve-fitting of spectra recorded with Mg K$\alpha$ source, meanwhile the spectra recorded with monochromatic Al K$\alpha$ were fitted in Avantage v5.9929 from ThermoScientific Co.  The backgrounds of inelastically scattered electrons were removed using the Shirley baseline model. For high-accuracy oxide thicknesses, the calculation was based on the procedure reported in the ISO 14701:2018 [26, 27]. For IMFP was calculated using the TPP2M formula contained in QUASES-IMFP-TPP2M Ver.4.1 software [28]. The information depth sampled, $d_\theta$, was calculated using the equation $d_\theta$= 3$\lambda$Sin($\theta$) [29].

**RESULTS AND DISCUSSION**

*Crystalline phase analysis and chemical composition distribution using HR-STEM/EDS*

X-ray diffraction is an essential and non-destructive technique to identify the phase composition and crystallinity degree in any solid material. Specifically, Grazing-Incidence X-ray diffraction (GI-XRD) is widely applied to the study of thin films, by controlling the depth penetration into the film by incidence angle or photon energies selection. The $Nb_3Sn$ samples under study were prepared using vapor diffusion technique with a mixture of $SnCl_2$ and Sn on an anodized Nb substrate; more details about sample preparation are available in the experimental section. The GI-XRD diffraction patterns for the $Nb_3Sn$ samples were recorded with a synchrotron photon energy of 15 keV and incident angle of 1˚. The diffraction peaks were indexed using the DicVol method and then verified by Le Bail profile fitting ( Figure 1a).  These XRD patterns reveal that the coating was successful and therefore the formed $Nb_3Sn$ has an ordered and regular structure. The unit cell found corresponds to a primitive cubic structure with Pm-3n space group and clearly shows a variation in the lattice parameter from 0.5308 to 0.5346 nm, probably related to a non-homogeneous local composition (discussed below). In the structure of the intermetallic material, Sn atoms are found forming a bcc sub-cell while the Nb atoms are situated on three orthogonal chains in the faces of a cube. Such chains are found parallel to the [100], [010], and [001] axes [30], Figure 1b. The Sn-Nb distance is about 2.95(9) Å and 2.64(10) Å for Nb-Nb, which is about ~7% smaller than the Nb-Nb distance in pure bcc-Nb. The high-resolution diffraction pattern reveals a mixture of fundamental peaks leading to a lattice parameter variation. The $Nb_3Sn$ fundamental peak splitting was previously observed by Becker et al. [31], where the fundamental peaks showed a shoulder on the high angular side. In that study, the high-resolution diffraction pattern was recorded with a $\theta$-$2\theta$ Bragg-Brentano approach. The pattern fitting reveals two $Nb_{1-\delta}Sn_\delta$ phases with $\delta$=18 and 25 %, with the former phase occupying 25% of the film volume. In our diffraction patterns, at the higher angular side recorded, e.g., plane (4 2 2), the fundamental peak splits into a set of components, Inset Figure 1a. Therefore, the lattice variation (0.5308 to 0.5346 nm) may result from a deviation in the Sn concentration, 18-25 % Sn, relative to the nominal stoichiometry $Nb_3Sn$, which crystallizes with a cubic unit cell [32]. Such a composition deviation is consistent with the $Nb_3Sn$ phase diagram. The presence of local defects such as vacancies also cannot be discarded. No reflections corresponding to Nb-substrate were identified in the recorded XRD pattern. This agrees with the estimated thickness, of about ~3 $\mu$m, for the deposited $Nb_3Sn$ film, while the calculated X-ray depth penetration in our experiment with wavelength $\lambda$= 0.826 nm and 1° degree corresponds to ~1 μm [23]. On the surface, the interatomic spacing is different from the bulk and therefore, no lattice order is found. Indeed, the diffraction patterns do not reveal the presence of any reflections of Nb-oxides. A TEM specimen was prepared using FIB, see Experimental section. Then, high-resolution Scanning Transmission Electron microscopy (HR-STEM) experiment was conducted. The HR-STEM image shows excellent agreement with XRD results with the absence of diffraction peaks from crystalline Nb-oxides. Figure 1d illustrates the amorphous nature of the oxide layer; its thickness was roughly estimated to be about 5±1 nm. Additionally, the micrograph reveals well-resolved $Nb_3Sn$ atomic columns, confirming the $Nb_3Sn$ structure. A drawing of the atomic ordering is illustrated in the Inset of Figure 1d. From the thin TEM specimen, the chemical composition was estimated with Energy Dispersive Spectroscopy (EDS) mapping. The elemental EDS maps reveal a layered distribution of Nb K, Sn L and O K, where the top surface is Sn-enriched (Figure 1c), followed by a saturated layer of Nb and O. This feature in the last spectra suggests that Nb-oxides are present on the Nb and Sn layer corresponding to $Nb_3Sn$. All these features agree with the HR-STEM image. Additionally, to corroborate the depth distribution observed by EDS mapping, specifically the composition of the topmost surface, complementary data were obtained by ARXPS measurements. Nb 3d, O 1s and Sn 3d spectra were recorded at four angles from 15 to 60 ˚ degrees, and the corresponding depth penetration was calculated (Fig. S3). All the spectra recorded were quantified using the sensitivity factors reported by Wagner and Scofield integrated in the Avantage software. As take-off angles increase, the atomic Nb:Sn ratio decreases from 2.3 to 1.9, revealing an increase of Sn at the surface. On the other hand, the Nb:O ratio increases from 0.37 to 0.45 atomic ratio as take-off angles decrease, the decrease of O to Nb ratio at deeper angles leads to the conclusion of higher near-surface O. Sn:O ratio at different angles does not show any definitive variation.

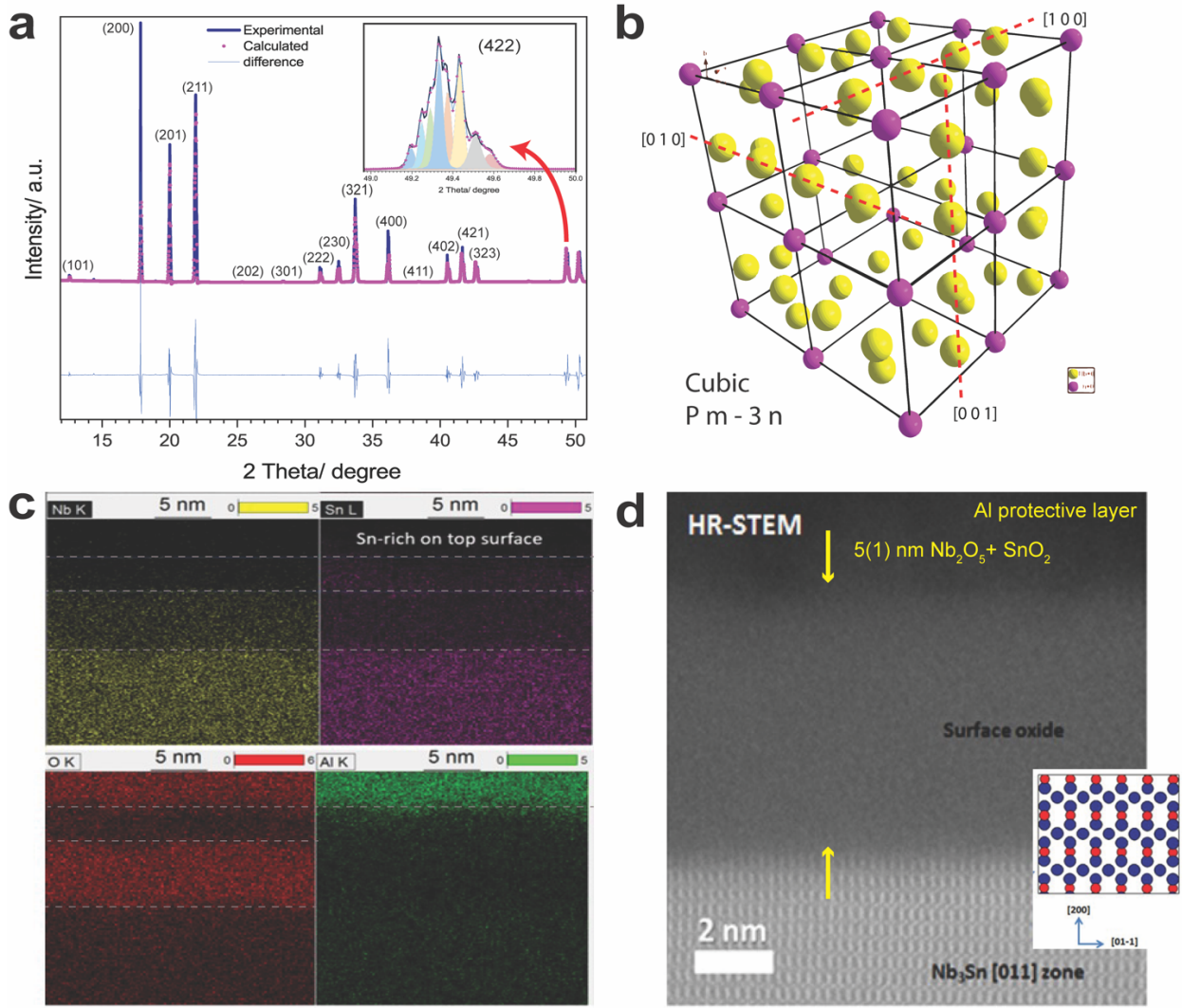

**Figure 1.** a) Nb$_3$Sn experimental and calculated XRD pattern recorded at a shallow angle of 1.°, b) Atoms packing within the Nb$_3$Sn framework, c) Elemental composition by EDS mapping, and d) high-resolution STEM image showing the amorphous oxide layer and crystalline and well-ordered Nb$_3$Sn atomic structure in the [0 1 1] direction. Inset: atomic model illustrating two atom sites: Nb (red) and Sn (blue).

*- Estimated thickness of the oxide layer from TEM and AR-XPS data*

It is important to identify oxide composition as well as the thickness of the natural oxide layer in EP-Nb and Nb$_3$Sn to better understand the role of the oxide layer in the low-field Q-slope [6]. Previous studies reported on the complex Nb-oxides on the surface of SRF Nb cavities after different chemical and heat treatments [6]. Delheusy et. al. reported from their studies [33], using photon energy variation XPS, that the topmost surface layer on Nb has 2-5 nm of niobium oxides. The Nb-oxide layer consists of $Nb_2O_5$ on top, followed by $NbO_2$, NbO, and metallic Nb. Additionally, two components identified such as P1 and P2 were fitted in the proposed model. P2 component was assigned to be $Nb_2O$, located below NbO, meanwhile P1 was assigned $Nb_2O_3$, found below P2, next to metallic niobium at the metal/oxide interface. In our study to determine the layered oxide ordering in the EP-Nb and $Nb_3Sn$ samples, we use ARXPS in an XPS lab-based instrument equipped with an AlK$\alpha$ source. As described in the experiment section, we use a conventional ARXPS approach [34] where the sample is tilted to a set of angles (15º, 25º, 35º, 45º, 60º, 80º) towards the energy analyzer opening. Besides, to calculate the thickness, the methodology according to ISO-14701:2018 was used [26]. The Nb 3d core-level appears as two peaks due to the spin-orbit effect. The Nb 3d curve-fitting model consists of two asymmetric peaks corresponding to metallic Nb, Nb $3d_{5/2}$ at 202.5 eV and the Nb $3d_{3/2}$ peak being at a 2.7 eV higher binding energy, with an area ratio of 3:2. Then two symmetric doublets from the $Nb_2O_5$ and $NbO_2$ oxides were fitted at 208 and 206.4 eV, with a spin-orbit difference of 2.74 eV and, each Nb $3d_{3/2}$ peak area being the 60 % of their $Nb_{5/2}$ counterpart. In a row, the contribution of Nb oxide with metallic nature was added by one doublet asymmetric peaks at 203.4 eV, and binding energy difference of 2.72 eV. Considering this curve-fitting model proposed for the sample EP-Nb, the calculated thickness using the equation (S1) and (S2) is estimated at $d_{Nb2O5}$= 4.9 nm. The Nb-suboxides (NbO and $NbO_2$) are estimated with lower accuracy resulting in $d_{NbO}$= 0.9 nm. Both thickness values then give the total oxide thickness of EP-Nb, which results in $d_{total}$= 5.8 nm. The same procedure was applied to calculate the effective oxide thickness for the $Nb_3Sn$ sample. The complex oxide layer in $Nb_3Sn$ consists of Nb and Sn oxides. Therefore, Sn 3d spectra were considered and fitted into two contributions, two asymmetric peaks with a binding energy difference between them of 8.4 eV and an area ratio of 3:2 corresponding to metallic Sn. The second set of symmetric peaks corresponding to Sn oxides, separated by 8.4 eV, with a 3:2 area ratio. The calculated thickness from Sn oxides is $d_{SnO2}$= 1.8 nm. The calculated thickness of niobium oxides in $Nb_3Sn$ is $d_{Nb2O5}$= 4.7 nm, followed by Nb suboxides with thickness $d_{NbO}$= 2.7 nm. As we describe above, the effective oxide thickness is the calculation of all the components' thickness ($d_{Nb2O5(1)}$ + $d_{Nb-sub}$), therefore the thickness of the oxide layer in $Nb_3Sn$ results in $d_{total}$= 7.4 nm. From the angle variation measurements, we suggest that the surface composition order is given as follows: on top, $SnO_2$ and $Nb_2O_5$ oxides are the intermixed species followed by the sub-oxides layer, and the metallic Sn and Nb components. $Nb_3Sn$ shows a more complex surface oxide, where $Nb_2O_5$ layer is about 0.2 nm thinner than in pure Nb, but niobium suboxides layer is 1.8 nm thicker, Figure S3 [inset]. These findings corroborate our early argument that the electron density donation of the Nb atoms towards Sn atoms leads to a stronger bond. In addition, shorter Nb-Nb distance gives rise to a stronger metallic bond, and, therefore, Nb reactivity is diminished to form bonds with other elements such as oxygen.

*-Thermal stability of Nb oxide layer in EP-Nb cutout*

The native oxide layer in pure-Nb has been characterized in the previous sections by a series of analytical techniques to determine the chemical and depth composition, amorphous nature, and thickness. In summary, we found that the passive oxide layer in the EP-Nb sample consists of disordered phases of $Nb_2O_5$ at the topmost surface with a thickness of 4.9 nm, followed by an Nb-suboxides layer of c.a. 0.9 nm, which is mainly formed by $NbO_2$ and $NbO$ phases. The removal of the native oxide layer in SRF Nb cavities enhances the quality factor $Q_0$ by decreasing the residual surface resistance [35]. These "oxide-free" SRF cavities undergo heat treatments above 300 ºC for a couple of hours and, remain under vacuum until cryogenic RF test to avoid a re-growth of a new oxide layer in air.

In this section, we analyze the thermal stability of the passive oxide layer in the EP-Nb sample up to 600 ºC. Unlike previous reports where the per-default spectrometer angle is used (commonly 45-60º), in our experiment the sample was tilted to 34º to access finer surface details by enhancing the surface sensitivity. The sample remains at this angle upon heating from room temperature to 600 ºC. The progressive dissolution of the native oxide layer was followed up by recording the Nb 3d, O 1s and C 1s sub-spectra every 25 ºC. At 200 ºC, the Nb 3d XPS spectrum, Figure 2, is fitted with four components: $Nb_2O_5$ at 208.0 eV (84.5 at. %), $NbO_2$ at 206.4 eV (10.6 at. %), $NbO$ at 203.4 eV (1.0 at. %), and metallic Nb at 202.5 eV (3.9 at. %). Upon heating at 225 ºC, the doublet peak assigned to $Nb_2O_5$ becomes broader along with a decrease in intensity. This doublet is then fitted into two phases, the original $Nb_2O_5$ and a new non-stoichiometric phase assigned to $Nb_2O_{5-\delta}$. Both phases represent the 74.1 at. % of total Nb. Nb 3d peaks of the $Nb_2O_{5-\delta}$ phase are shifted 0.5 eV towards lower binding energy as compared to Nb 3d peaks for the $Nb_2O_5$ phase. Upon heating, Nb-O bond breakage within the amorphous $Nb_2O_5$ network, where Nb atoms are surrounded by oxygen atoms with electronegative character, causes the re-distribution of electron density around Nb atoms and the corresponding 0.5 eV chemical shift in the binding energy of Nb 3d electrons. Next to $Nb_2O_{5-\delta}$ peak, the contribution of $NbO_2$ is present at 206.4 eV with 12.1 at. % of total Nb. Metallic Nb and NbO represent the 4.7 and 1.1 at. % of total Nb. Upon heating to 325 ºC, the Nb 3d spectra can be decomposed without $Nb_2O_5$ oxide contribution suggesting the complete dissolution of this oxide. On the other hand, $Nb_2O_{5-\delta}$ phase is reduced to 2.3 at. %, $NbO_2$ to 5.7 %, and most of the oxides are reduced to NbO and metallic Nb with 53.4 at. % and 9.7 at. %, respectively. At 425 ºC, $Nb_2O_{5-\delta}$ phase is finally completely dissolved. The Nb 3d spectrum shows the existence of $NbO_2$ (206.5 eV), NbO (203.5 eV), and metallic Nb (202.5 eV) with atomic percentages of 5.5, 38.9 and 9.3 of total Nb. At 525 ºC, $NbO_2$ is still present with 5.8 at. % at 206.5 eV as binding energy (Nb $3d_{5/2}$). Metallic Nb represents the 5.4 at. % of total Nb. Then, the main composition at 525 ºC is NbO with 80.4 at. %. The peak corresponding to NbO is shifted 0.3 eV towards the high binding energy side. Eventually, at 600 ºC, the maximum temperature reached in this experiment, the Nb 3d spectrum is composed of metallic Nb at 202.5 eV (4.2 %), NbO at 203.7 eV (89.2%), NbC at 204.6 eV (2.8 %) and $NbO_2$ at 206.5 eV (3.8 %).

Besides Nb oxide evolution upon heating, Nb-C and Nb-O-C phase formation is found above 225 ºC. At 225 ºC the simultaneous formation of two new intermedia phases [36] is observed, corresponding to $NbO_xC_y$ and Nb-carbides. The phase identified such as $NbO_xC_y$ and NbC represents 6.0 and 1.8 % of total Nb, with binding energies of 205.6 and 204.5 eV, respectively. The formation of Nb-carbides [37] could be the result of the diffusion process of C into the near-surface and grain boundaries upon heating. The source of C is related to the hydrocarbons (C-H, C-C, C=C, etc.) that exists on the surface, due to air exposure and handling of samples before analysis. To confirm the presence of Nb-carbides proposed in our Nb 3d curve-fitting model, the C 1s spectra were also analyzed upon heating from room temperature to 600 ºC, Figure

S7. A new peak at 283 eV along with the C adventitious at 284.6 eV is observed in the C 1s spectrum at 225 ºC. This phase was monitored by the C 1s spectrum and the synthetic peak in Nb 3d at 204.4 eV showing high stability up to 600 ºC. On the other hand, the intermedia phase, $NbO_xC_y$, exists only at 225 ºC and could be the precursor for the Nb-carbides phase. The presence of $Nb_2O_3$ is also observed at 325 and 425 ºC and then it is reduced to NbO. Nb 3d spectra analysis of the dissociation of the oxide layer is complemented with the O 1s spectrum curve-fitting, where the oxygen species were identified, see Figure S5.

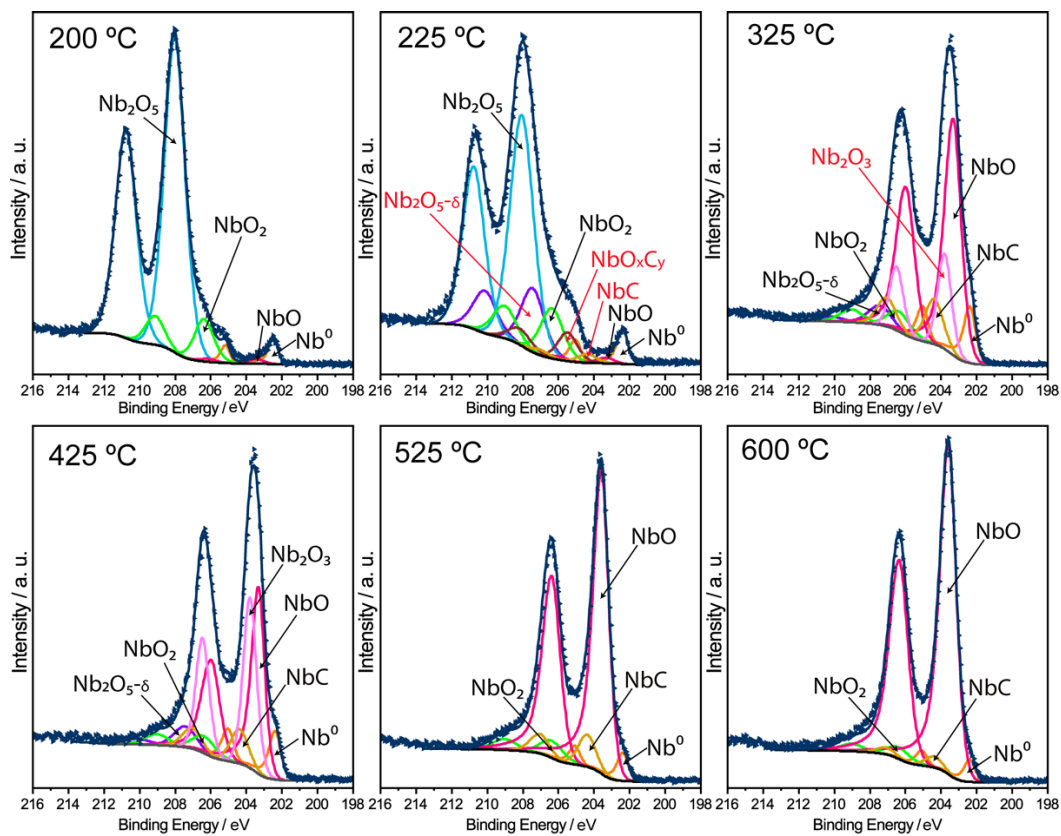

**Figure 2.** Nb 3d species XPS curve-fitting resulting for the EP-Nb sample upon heating to 600 ºC. The EP-Nb sample was extracted from a 1.3 GHz SRF cavity, baseline-treated. Sample treatment is described elsewhere [6].

*-Thermal stability of Nb₃Sn oxide layer at high temperatures (> 600 ℃)*

Post-coating heat treatments are the potential post-processing technique for $Nb_3Sn$ SRF cavities to improve their performance. As mentioned above, heat treatments applied to pure-Nb SRF cavities aim to reduce residual resistance through the dissolution of the oxide layer and to improve BCS resistance through impurity diffusion. In our study, we characterized the $Nb_3Sn$ oxide layer and its evolution upon heating by EDS mapping, HR-STEM, and ARXPS to determine the depth chemical distribution, crystalline order, and thickness. As previously discussed, the complex amorphous layer in $Nb_3Sn$ consists of 4.7 nm of $Nb_2O_5$ intermixed with $SnO_2$ on the topmost surface and 2.7 nm of Nb-suboxides. $Nb_2O_5$ layer in $Nb_3Sn$ is thinner than it is in Nb due to a lower reactivity of Nb atoms, surrounded by Sn atoms, to oxygen. We suggest that the strong overlap of the Nb and Sn electronic structure leads to a higher thermal stability of $Nb_3Sn$ oxide layer. We have not found Nb – Sn – O phase diagram that could help us evaluate the thermal stability of $Nb_3Sn$ oxide layer. In previous *in-situ* XPS studies performed by this group and unpublished, $Nb_3Sn$ sample was *in-situ* heated under UHV conditions at ~300 ºC for 1 h, and, unexpectedly, the peaks corresponding to $Nb_2O_5$ phase were still present, revealing a higher stability of $Nb_2O_5$ in $Nb_3Sn$ than in pure-Nb. The $Nb_2O_5$ peaks are no longer observed in the Nb 3d spectrum recorded from Nb upon ~300 ºC for 1 h, while in $Nb_3Sn$, the $Nb_2O_5$ peaks are dominating the spectrum. In the following section, we followed up on the evolution of the oxide layer in $Nb_3Sn$ from 200 to 700 ºC to determine the temperature of dissolution of the oxide layer as well as to identify the phases resulting from the reduction of oxide phases. This study is the first of its kind to study the nature of $Nb_3Sn$ oxide layer along with its thermal stability in order to develop post-coating heat treatments on $Nb_3Sn$ SRF cavities to decrease the surface resistance at low temperatures.

*Sn MNN Auger complemented with Sn 3d*

After discussing the intermixed composition of the oxide layer in $Nb_3Sn$, which consists of $Nb_2O_5$ and $SnO_2$, we will now discuss the Sn phases´ evolution upon heating from room temperature to 600 ºC. The poor sensitivity of the Sn 3d photoelectron signal in the identification of the oxidation states ($4^+$, $2^+$ and metallic $Sn^0$) of Sn is well known [38, 39]. Therefore, we complemented the analysis of Sn 3d spectra, with the analysis of the Auger signal, Sn MNN, where profile features accurately reveal the corresponding valence [40]. At room temperature, the Sn 3d photoelectron spectrum consists of two peaks: Sn $3d_{5/2}$ and Sn $3d_{3/2}$ with an area ratio of 3:2. At 486.8 eV, a symmetric peak assigned to the $SnO_2$ phase is found with 91.4 atomic %. On the lower binding energy spectral region, an asymmetric peak was fitted with a binding energy of 484.5 eV and 8.6 atomic % that corresponds to metallic Sn, Figure S4 a. Upon progressive heating to 200 ºC, the $SnO_2$ peak area decreases and the ratio changes to 2.8: 1 ($Sn^0$). A progressive evolution was observed at 325 ºC in the peak areas, see Figure S4 b, where 39% of Sn dioxide is reduced to metallic Sn, and therefore $SnO_2$ peak is observed as a small shoulder on the higher binding energy side of $Sn^0$. Further heating 425 ºC, a very-weak peak just above the background appears at 486.8 eV which corresponds to 8% of total Sn meanwhile the 92% is metallic Sn (484.5 eV). The absence of any weak shoulders next to the metallic Sn peak at 525 ºC indicates the complete dissolution of Sn dioxide. The Sn 3d spectra recorded at 525 and 700 ºC only show an asymmetric peak at 484.5 eV (Sn $3d_{3/2}$), and its Sn $3d_{5/2}$ spin-orbit component is 8.4 eV above. Figure S4 c shows the XAES Sn MNN lines recorded in the interval from 200 to 600 ºC. In principle, the Auger spectrum at 200 ºC consists of two broad peaks at kinetic energies of 425 and 433 eV. Both signals were assigned to oxidized Sn, meanwhile $Sn^0$ is observed as small peaks at higher kinetic energies of 437.2 eV and 428.7. Upon heating, most oxides are reduced to metallic Sn. Therefore, the following Auger spectra from 425 to 700 ºC are composed of a fine structure corresponding to two main peaks $^1G_4$ and their respective multiplets, which result from the $4d^8$ final-state configuration [41]. The $^1G_4$ spin-orbit splitting in our spectra with an energy difference of 8.5± 0.1 eV agrees with the literature, and therefore corroborates the complete reduction of Sn-oxides upon heating to 525 ºC.

*Nb 3d core-level region*

Building on Sn oxide dissolution discussion, we now turn to the niobium oxide evolution in $Nb_3Sn$ with the analysis of Nb 3d spectrum. All the spectra were taken at the shallow angle of 34 º upon heating. Nb 3d spectrum at 200 ºC consists of two doublets peaks assigned to metallic Nb at 202.8 eV (Nb $3d_{5/2}$) and $Nb_2O_5$ at 208.1 eV (Nb $3d_{5/2}$), with an area ratio between them 1:3. We applied the same curve-fitting model used for Nb 3d in the EP-Nb sample. Although in $Nb_3Sn$, the spectrum at 200 ºC, $Nb_2O_5$ at 208.1 eV occupies only 8.9 atomic % of total Nb, suggesting an early reduction of the original $Nb_2O_5$ into the non-stoichiometric $Nb_2O_{5-\delta}$ at 207.9 eV(50.3 at. %). The rest of the fitted phases are $NbO_2$ at 206.4 eV (6.6%) and NbO at 203.4 eV (6.3%) and Nb-Sn at 202.8 eV (27.9%). Upon heating to 225 ºC, two new phases appear. The simultaneous reduction of $NbO_2$ leads to the formation of $NbC_xO_y$ phase at 205.6 eV as well as $Nb_xC_y$ (Nb-carbides) at 204.5 eV. The curve-fitting parameters and phase identification are available in the supplementary information file, Table S2. Upon heating to 325 ºC, the spectrum profile does not show major modifications in terms of binding energy values for $NbO_2$, NbO, $Nb^0$, NbC, $Nb_2O_{5-\delta}$ and $Nb_2O_5$. Nevertheless, the atomic percentage of $Nb_2O_{5-\delta}$ decreases from 50.3 at.% (200 ºC) to 43.7 at. %, and simultaneously the contribution of $NbO_2$, NbC and $NbC_xO_y$ increases to 7.2, 2.8, and 4.3 at. %. The continuous reduction of $Nb_2O_{5-\delta}$ progresses slowly at 425 ºC. The double peaks assigned to $Nb_2O_{5-\delta}$ represents the 35.1 at. %, and Nb-Sn represents 17.1 at. % of total Nb, meanwhile the rest of Nb is shared by the phases of $NbO_2$, NbC and $NbC_xO_y$. Upon heating to 525 ºC, the $Nb_2O_{5-\delta}$ phase at 207.7 eV and $Nb_2O_5$ phase at 208.1 eV peaks decrease in intensity and the calculated atomic percentage is 29.2 % of total Nb. Along with the reduction of $Nb^{5+}$-containing phases, NbO, NbC, and $NbC_xO_y$ increase their atomic percentage to 43.2 %. The decrease for the $Nb^{5+}$ phases is reflected in a distortion of the spectral profile as observed in Figure 3, where the gap between the Nb 3d spin-orbit peaks is no longer resolved. This feature is mainly due to the increase in the $NbO_2$ and NbO phases. No changes in the binding energies are observed. Consequently, for the Nb 3d spectrum curve-fitting, recorded at 600 ºC, the binding energy values assigned for each phase were fixed, and only the intensity was allowed to vary according to the experimental spectral profile. At this temperature, a new phase appears, and it is assigned to $Nb_2O_3$ at 204.0 eV which represents the 7.9 at. %. Although, all the phases are present, the total intensity of $Nb_2O_{5-\delta}$ at 207.6 eV and $Nb_2O_5$ at 208.1 eV decreases to 21.8 at. %. From these experiments, we find that $Nb^{5+}$ phases ($Nb_2O_{5-\delta}$ and $Nb_2O_{5)}$ are present even upon heating to 600 ºC.

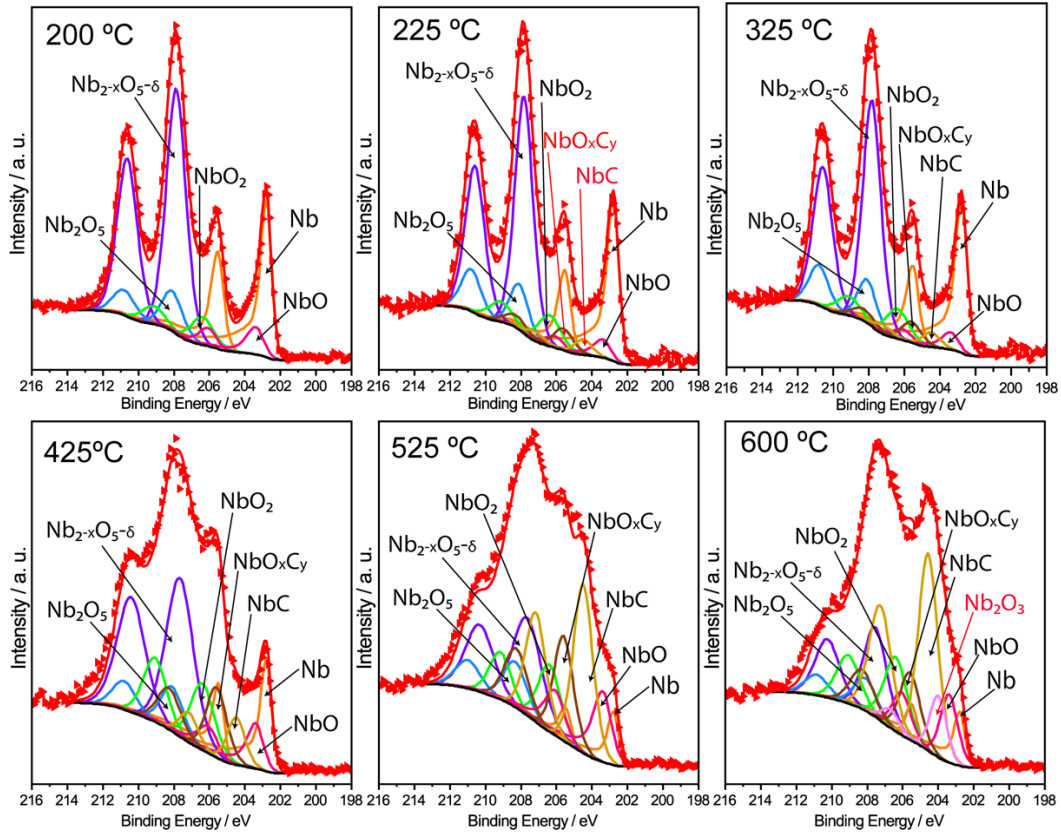

**Figure 3.** Nb 3d XPS curve-fitting resulting from Nb₃Sn-coated SRF grade Nb sample, clearly showing the high stability of the Nb-suboxides at 600 ℃.

The next set of measurements was carried out in XPS instrument equipped with a heating stage that can reach the maximum temperature of ~1300 ℃ under ultra-high vacuum environment. Contrary to the previous setup, where the angle was modified at 34º for most surface sensitivity, all the spectra were recorded under the normal "bulk" angle upon heating from room temperature to 700 ℃. Figure 4 shows the progressive evolution of the Nb₃Sn oxide layer with the curve-fitting of Nb 3d spectrum upon heating from 325 to 700 ℃. As already discussed, the double peaks corresponding to Nb₂O₅ and Nb₂O₅₋δ are observed at temperatures of 325 and 425 ℃ with an atomic percentage decrease of 68.9 to 52.3. Like the earlier experiment at 525 ℃, the Nb 3d spectrum is well-fitted into the four phases assigned to NbO₂, NbC, NbCₓOᵧ and metallic Nb. Besides these phases, peaks corresponding to Nb₂O₅ and Nb₂O₅₋δ were also defined with the peak intensity decreasing to 20.5 % and 1.9 %, respectively. There are differences in the spectra recorded at 600 ºC between the two experiments. In the second measurement, the Nb 3d spectrum was fitted only into six components assigned to $NbO_2$, NbC, $NbC_xO_y$, NbO and metallic Nb. Further, a very weak peak of $Nb_2O_{5-\delta}$ was observed with a binding energy at 207.7 eV. Unlike our first experiment, where both $Nb^{5+}$ phases were well-fitted with a presence of 21.8 atomic percentage at 600 ºC, in this new measurement, no traces of original $Nb_2O_5$ phase is observed. On that ground, we suggest the dissolution of the original $Nb_2O_5$ oxide layer occurs close to 600 ºC, but the reduced $Nb_2O_{5-\delta}$ phase is still present with a concentration of 20.5 atomic % of total Nb. The difference in the results between two experiments can be explained by the shallower detecting angle used in the first experiment, which results in a more surface sensitive capture of the Nb 3d and O 1s spectra that could be hidden in the normal "bulk" measurements. Subsequently to the reduction of the $Nb^{5+}$ containing phases, the spectra is composed by $NbO_2$ (4.6 %), NbC (5.6 %), $NbC_xO_y$ (13.7 %), NbO (35.7 %) and Nb-Sn (36.3 %). Upon heating to 625 ºC, the complete dissolution of $Nb_2O_{5-\delta}$ and $NbC_xO_y$ takes place. The Nb 3d spectra is then composed of the remaining four phases, $NbO_2$ (4.5 %), , NbC (13.2 %), NbO (33.3 %) and Nb-Sn (49 %). We observed that, upon heating from 655 to 700 ºC, recombination of the available oxygen atoms in the near surface with Nb take place. Such a progressive process is observed as an increase in the atomic concentration of NbO accompanied by a decrease of Nb.

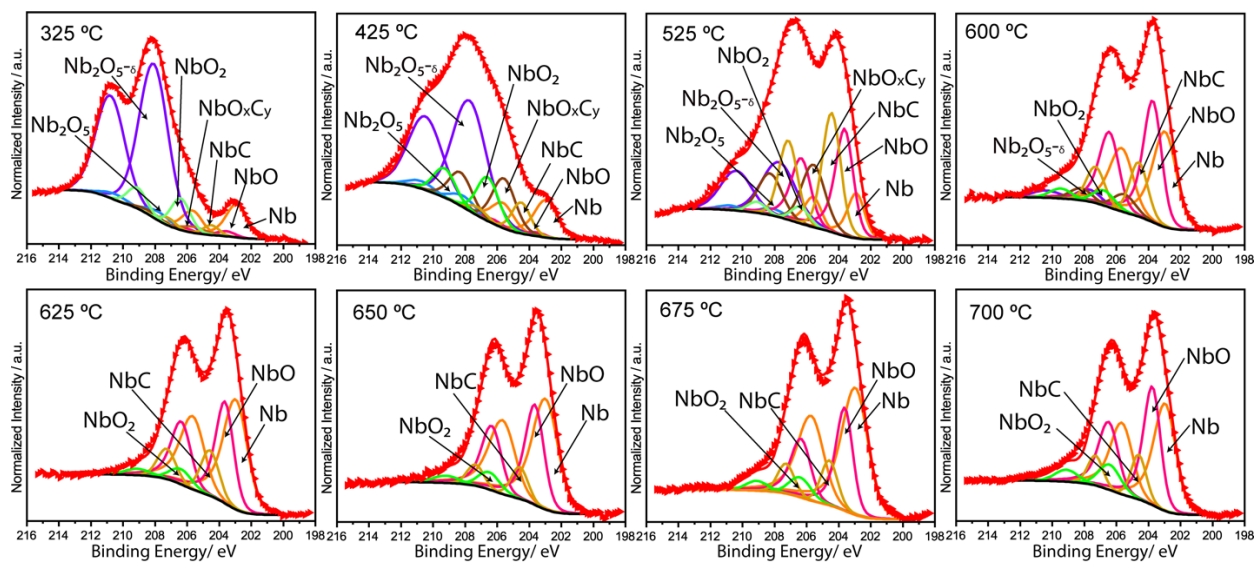

**Figure 4.** XPS Nb 3d curve-fitting: progressive evolution of the surface oxide layer in Nb$_3$Sn-coated SRF grade Nb, upon heating from 325 ºC to high temperature ( ~700 ºC).

*-Carbonaceous surface-contamination removal from 145 to 200 ºC*

In the EP-Nb cut-out analysis upon heating, the vacuum pressure abruptly increased in the range from 145 to 200 ºC. To maintain the base pressure in the analysis chamber at $10^{-9}$ mBar, we reduced the temperature ramp rate and started the spectra acquisition every 10 ºC. The degassing stage, as we are going to refer to the range from 145 ºC to 200 ºC, was analyzed and quantified from O 1s and C 1s spectra. Figure S7 shows the decomposition of the carbonaceous phases observed as a broad shoulder at 286.2 eV. This signal disappears upon heating to 155 ºC, and it is accompanied by a decrease in the atomic percentage of oxygen. This is followed by the second step in the two-step process: the evolution of the CO-contained compounds and O-contained species from 155 to 165 ºC. Therefore, we could assume that the degassing stage in EP-Nb cut-out is characterized by the evolution of C and O-contained compounds, mainly located at the surface. Once overcome the critical range from 145 to 165 ºC, the atomic percentage of C remains stable up to 200 ºC. We suggest that a fraction of C and O atoms diffuse toward the near-surface as observed by the formation of NbO$_x$C$_y$ and Nb-carbides phases from the Nb 3d and C1s spectra. Meanwhile, upon heating of Nb$_3$Sn sample, the vacuum pressure remained stable at $10^{-9}$ mBar up to 700 ºC. We didn't register any anomaly in the critical range from 145 to 165 ºC which could be related to the desorption of carbonaceous contaminants. A second measurement with an emphasis on the temperature range from 150 to 165 ºC was carried out in another XPS spectrometer. Contrary to the C 1s spectra recorded from EP-Nb, the C 1s spectra recorded from Nb$_3$Sn consists of one symmetric peak, with no evidence of shoulders, see Figure 7 b). The vacuum pressure during this experiment remains stable at the critical temperature range, 150 to 165 ºC, and no evidence of changes in the atomic percentage of C was observed. Compared to Nb metal, these experiments suggest that Nb$_3$Sn may have a low affinity for the absorption of carbon-based phases, and it was evident from the recorded XPS spectra upon heating below 300 ºC that these phases migrate to the outer most surface region as detected and quantify in the recorded XPS spectra. Like oxygen, carbon is an electronegative element, and a lower electron density of the Nb atom in the presence of Sn may result in a lesser affinity to the carbon-based phases absorption.

*-On the nature of the thermal stability in the native oxide layer for pure Nb and* $Nb_3Sn$

Figure 5a shows the XPS core levels of pure-Nb and $Nb_3Sn$ samples, with emphasis on the Nb phases. Both, $Nb_{5/2}$ and $Nb_{3/2}$ peaks in $Nb_3Sn$ are found to be shifted by 0.3 eV to a higher binding energy relative to the analog peaks in Nb metal. The binding energy value probes the electron density on the involved atom. An increase in the electron density results in a weaker interaction of the nuclear charge with their electrons and it is probed as a lower value for the binding energy. Inversely, when electron density is removed from the atom through an oxidation process, or when electrons are donated to other atoms during the formation of a chemical bond, the binding energy value can sense such an effect as a higher value for the binding energy. This is congruent with the dependence shows in Figure 5b for the value of the binding energy for different oxidation states of Nb [42]. In fact, the binding energy shift for Nb 3d in $Nb_3Sn$ relative to Nb metal is a definite clue on the charge donation to the neighboring Sn atoms in the intermetallic compound. The difference corresponds to a lower electron density on the Nb atom in the intermetallic solid as the consequence of Nb atoms interaction with their Sn neighbors in the solid structure. This must be observed as a lower binding energy for $Sn^0$ species, a higher electron density on the Sn atom. Such an expected shift is certainly observed as a lower value of binding energy for Sn in $Nb_3Sn$ relative to the one observed for Sn metal (Figure 5c). This is congruent with the observed for the Sn MNN Auger region. When that region was examined for XPS spectra corresponding to metallic Sn and $Nb_3Sn$, the Sn MNN Auger peaks shown a definite energy shift relative to the one observed for $Sn^0$, which is characteristic of a higher electron density on the tin atom in the intermetallic compound (Figure 5d). For comparison, the Auger spectral region for $SnO_2$ was included. In this tin oxide the tin atom has a lower electron density, and its Auger peaks appear with a pronounced shift toward to higher values of kinetic energy.

When the recorded Nb XPS sub-spectra for Nb metal and $Nb_3Sn$ were re-examined, such a binding energy shift is observed for the considered temperature region, where the probed surface region in the intermetallic compound remains stable. This finding supports the initial hypothesis regarding the nature for the observed differences between the surface oxide layer in Nb metal and $Nb_3Sn$ in terms of thickness, composition, changes on heating, and thermal stability. A similar effect is expected for other Nb-containing intermetallic compounds with elements of different electronegativity, e.g., Nb-Sb, Nb-Ir (Figure 5b). The difference in electronegativity for the involved metals senses their bonding interactions and the related electron density donation/subtraction effects. The Nb and Sn 3d spectral regions shown in Figure 5a and 5c appear free of the corresponding oxide phases because the oxide layer was removed through an Argon ion beam to obtain the spectra of the naked metal surface.

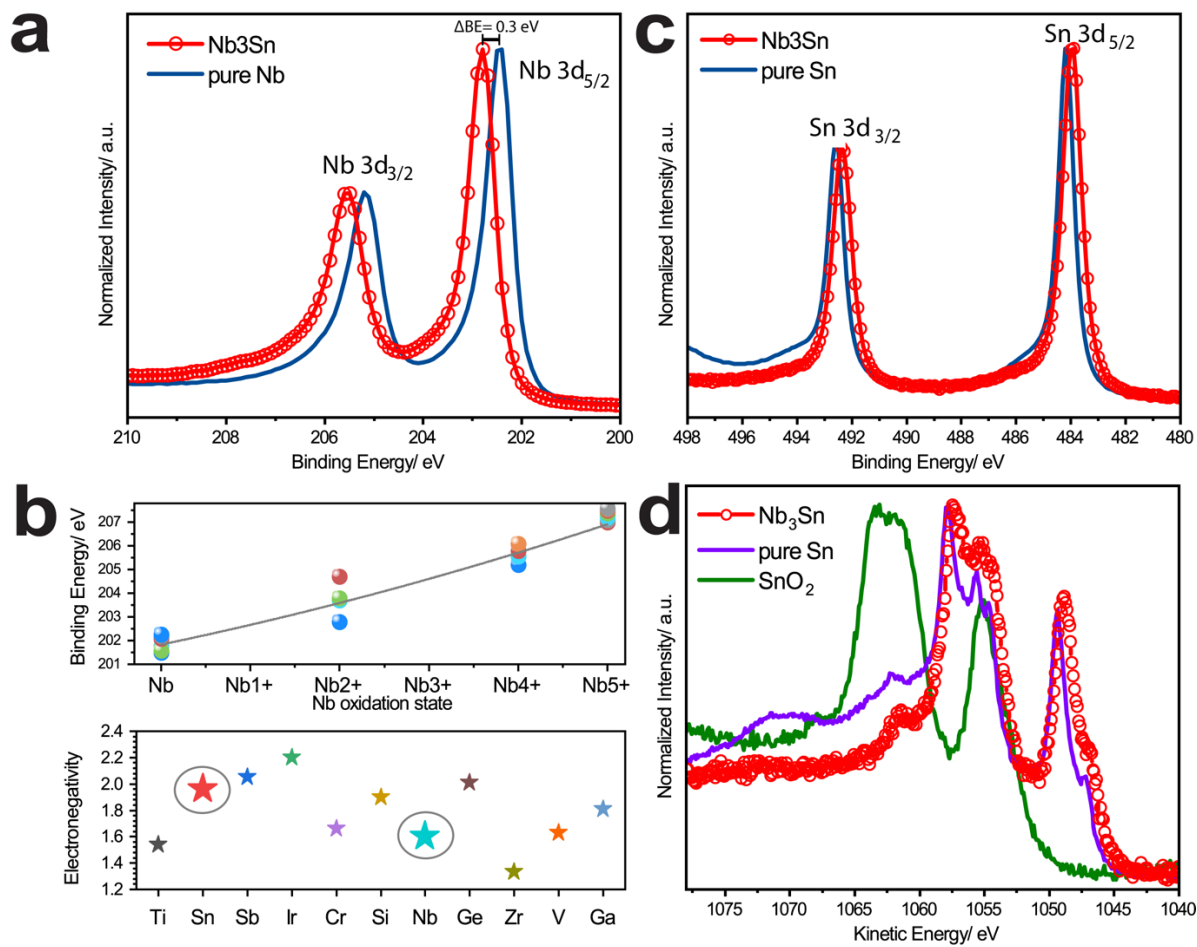

**Figure 5.** Chemical shift observed in the a) Nb 3d XPS spectra recorded from pure-Nb and $Nb_3Sn$, b) Sn 3d and d) Sn MNN feature Auger lines. On the bottom right c) represents the dependency of binding energy

respect to the density electronic in an atom, as well as the differences on electronegativities of the involved atoms in $Nb_3Sn$. The Nb and Sn 3d spectral regions were recorded for the surface where the oxide layer was removed through a monoatomic Ar ion sputtering at 2500 eV for 90 seconds.

**CONCLUSIONS**

In summary, this contribution has compared the oxide surface layer for Nb metal and $Nb_3Sn$ in terms of the thickness, chemical and phase composition, and behavior on heating, including the thermally induced reduction processes. The surface oxide layers in these two materials show significant differences both in as-formed oxide layer and upon heating. Such differences are explained by the electronic structure of Nb atom in these two materials and its reactivity toward the oxygen atom to form oxides. The interaction of Nb with Sn in $Nb_3Sn$ intermetallic solid involves charge donation from Nb atoms to the neighboring Sn atoms. This is effectively seen in XPS as a lower electron density on the Nb atom in $Nb_3Sn$ relative to the value obtained for metallic Nb. The electron density is sensed through the energy shift in the core-level binding energy. Such a difference between the two materials was observed in a wide range of temperatures. These results, on the nature of the observed differences between Nb and $Nb_3Sn$ materials, are relevant to the manufacture and processing of $Nb_3Sn$-based devices that operates in the superconducting state, and also guide the engineering processes towards improved performance of Nb metal-based devices, such as SRF cavities for particles accelerators and quantum computing technologies, among others.

**ASSOCIATED CONTENT**

**Supporting information.**

The Supporting Information is available free of charge on the XXXXX at DOI: XXXXXXX

**AUTHOR INFORMATION**

**Corresponding Authors**

*E-mail: grigory@fnal.gov (G. V. Eremeev)

*E-mail: sposen@fnal.gov (S. Posen)

**ORCID**

Arely Cano: 0000-0003-3743-4411

Juan Rubio Zuazo: 0000-0003-0614-5334

## AUTHOR CONTRIBUTIONS

A.C. and S.P. planned the research, proposed, and submitted the project; A.C. and B.L. carried out the *in-situ* XPS lab-based measurements; J-Y. L. carried out the high-resolution EDS and STEM experiments; J.R.Z. performed the high-T XPS measurements; G.E., M. M., A.R. and S.P. supervised the project. All authors interpreted and discussed the results and contributed to editing the manuscript.

## COMPETING INTERESTS

The authors declare no competing interests.

## ACKNOWLEDGEMENTS

This manuscript has been authored by Fermi Research Alliance, LLC under Contract No. DE-AC02-07CH11359 with the U.S. Department of Energy, Office of Science, Office of High Energy Physics. Award Number ECCS-2025124. The experimental study present herein was granted partial financial support by The Explore Nano Program from the Minnesota Nano Center. The authors are grateful to the SpLine staff for their valuable help and to the financial support from the Spanish MINECO and Consejo Superior de Investigaciones Cientificas under Grant NoMAT2011-23785, 2010 6 OE 013, ICTS-2007-02, MAT2005-25519-E, 2004 5 0E 292 andMAT2002-10806-E. A. Cano thanks to J.R. Zweibohmer for administrative and support services.